\documentclass[showpacs,preprintnumbers,amsmath,amssymb,aps,superscriptaddress]{revtex4}

\usepackage{graphicx}
\usepackage{amssymb}
\usepackage{epstopdf}
\usepackage{bm}

\begin{document}
    
\title{Precision Measurement of the n-$^3$He Incoherent Scattering Length Using Neutron Interferometry}
    
\author{M.~G.~Huber}
\affiliation{Tulane University, New Orleans, LA 70118}
\author{M.~Arif}
\affiliation{National Institute of Standards and Technology, Gaithersburg, MD 20899}
\author{T.~C.~Black}
\affiliation{University of North Carolina at Wilmington, Wilmington, NC 28403}
\author{W.~C.~Chen}
\affiliation{National Institute of Standards and Technology, Gaithersburg, MD 20899}
\affiliation{Indiana University, Bloomington, IN 47408}
\author{T.~R.~Gentile} 
\affiliation{National Institute of Standards and Technology, Gaithersburg, MD 20899}
\author{D.~S.~Hussey}
\affiliation{National Institute of Standards and Technology, Gaithersburg, MD 20899}
\author{D.~A.~Pushin}
\affiliation{National Institute of Standards and Technology, Gaithersburg, MD 20899}
\author{F.~E.~Wietfeldt}
\affiliation{Tulane University, New Orleans, LA 70118}
\author{ L.~Yang}
\affiliation{National Institute of Standards and Technology, Gaithersburg, MD 20899}
\date{\today}

\begin{abstract}
We report the first measurement of the low-energy neutron-$^3$He incoherent scattering length using neutron
interferometry: $b_i' = (-2.512\pm 0.012\mbox{ statistical}\pm0.014\mbox{ systematic})$ fm. This is in good agreement with a recent
calculation using the AV18+3N potential. The neutron-$^3$He scattering lengths are important for testing
and developing nuclear potential models that include three nucleon forces, effective field theories for few-body nuclear systems,
and neutron scattering measurements of quantum excitations in liquid helium. This work demonstrates the first use of a polarized nuclear
target in a neutron interferometer.
\end{abstract}

\maketitle
Recent years have seen remarkable advances in the decades-long struggle to compute nuclear structure and low-energy observables from the nucleon-nucleon (NN) potential. Modern NN potentials such as AV18 \cite{Wir95}, CD-Bonn \cite{Mac01}, and the Nijmegen potentials \cite{Sto94} produce good fits to experimental NN scattering observables up to 350 MeV. Quantum Monte Carlo methods which include three-nucleon (3N) potentials such as UIX \cite{Pud97} precisely reproduce the level spectra of light nuclei up to $A=12$ \cite{Pie01a,Pie01b,Pie02}. Effective field theory methods, which are logically related to the QCD Lagrangian and allow for better estimates of theoretical uncertainty, show promise for similar success in the near future \cite{Bed02}. But even with the inclusion of many-body forces, modern NN potentials have had only modest success in
predicting scattering observables in light nuclei. The aim of our program is to contribute to a complete set of basic experimental results for light nuclei which can be used in formulating the next generation of NN and many-body potentials.
\par
Neutron scattering lengths at the zero-energy limit can be directly and accurately measured using neutron interferometry. These are excellent high-precision bench marks for testing and/or calibrating the latest few-nucleon theoretical methods. Recent precision scattering length measurements include $n$-H, $n$-D, and $n$-$^3$He (coherent) \cite{Sch03,Bla03,Huf04,Ket06}. In general the agreement between these data and the best theoretical models, including 3N forces, is not very good. The situation with $^3$He is further clouded by the fact that two recent precision determinations of the coherent scattering length disagree by more than seven standard deviations \cite{Huf04,Ket06}. The incoherent scattering length was recently measured from the pseudomagnetic neutron spin rotation in polarized $^3$He gas \cite{Zim02}. Here we report on the first direct measurement of the $n$-$^3$He incoherent scattering length using neutron interferometry. This is also the first successful use of a polarized nuclear target in a neutron interferometer. The incoherent scattering length is also important for interpreting experimental studies of the scattering law $S(Q,\omega)$ of liquid $^3$He, as pointed out in \cite{Zim02}.
\par
A neutron interferometer  \cite{Sears,KR,RW} splits the matter wave of a single neutron into two coherent paths and then recombines them using Bragg diffraction perfect single-crystal silicon. A target placed in one beam path of the neutron interferometer produces a phase shift $\phi = -N_3 \lambda b' z$
where $N_3$, $z$ are the nuclear density and length of the target ($^3$He in this case), $\lambda$ is the neutron deBroglie wavelength, and $b'$ is the real part of the bound neutron scattering length of the target atom. The bound scattering length is appropriate because there is no momentum transfer between the neutron and target. It is related to the free atom scattering length by $a = b A / (A + 1)$, with $A$ the atom/neutron mass ratio. In general the interaction amplitude is complex and spin-dependent so the scattering length depends on the nuclear spin {\boldmath $I$} and neutron spin {\boldmath $\sigma_n$}:
\begin{equation}
b = b' + ib'' = b_c + \frac{2 b_i}{\sqrt{I(I+1)}} \mbox{\boldmath $I \cdot \sigma_n$}
\end{equation}
with the coherent ($b_c$) and incoherent ($b_i$) scattering lengths defined by
\begin{equation}
b_c = \frac{I+1}{2I+1}b_+ + \frac{I}{2I+1}b_-
\end{equation}
\begin{equation}
\label{E:incoh}
b_i = \frac{\sqrt{I(I+1)}}{2I+1}(b_+ - b_-)
\end{equation}
and $b_+$ and $b_-$ correspond to total spin channels $I+\frac{1}{2}$ and $I-\frac{1}{2}$.  If two
phase shift measurements are made: $\phi^\uparrow$ where {\boldmath $\sigma_n$} and $P_3$ (the $^3$He polarization) are parallel, and $\phi^\downarrow$ where {\boldmath $\sigma_n$} is reversed by a neutron spin flipper, with $\Delta \phi = \phi^\uparrow - \phi^\downarrow$ we have
\begin{equation}
\label{E:maineq}
\Delta b' = b'_+ - b'_- = -\frac{2\Delta \phi}{N_3 \lambda z P_3}
\end{equation}
\par
\begin{figure}
\includegraphics[width=5.0in]{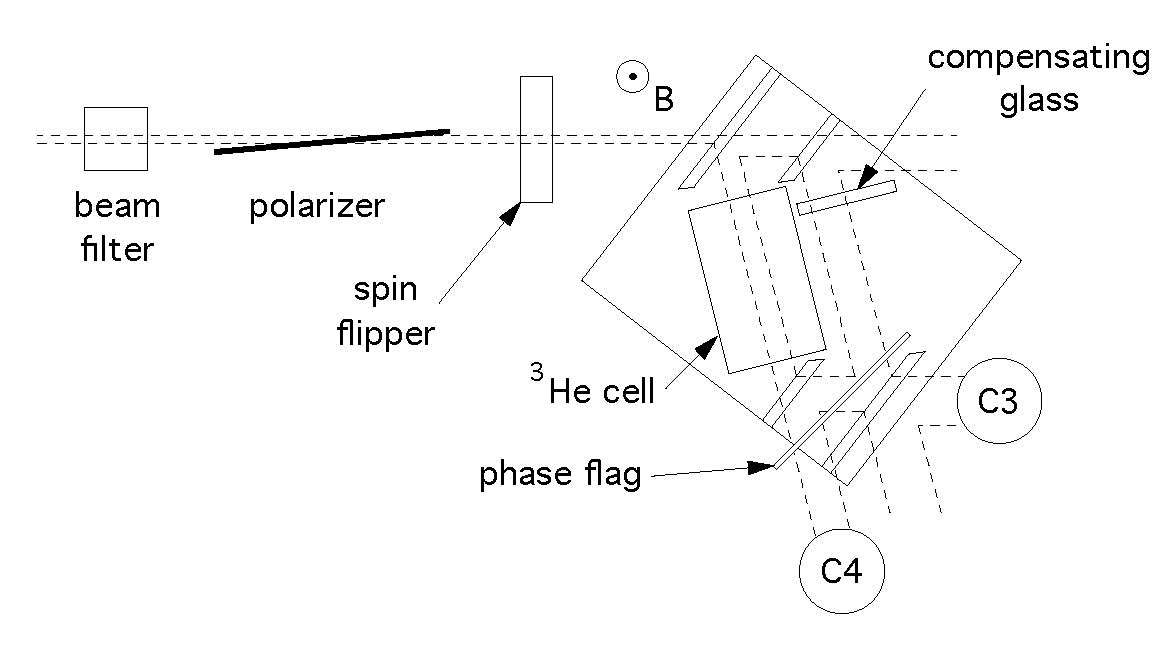}
\caption{\label{F:nHe3setup}Schematic representation of the experiment (not to scale). The dashed lines indicate the neutron beam paths. The
interferometer crystal was fabricated by the Physics Machine Shop at the University of Missouri, Columbia \cite{All98}.}
\end{figure}

The experiment was performed at the NIST Center for Neutron Research's Neutron Interferometry and Optics Facility \cite{NIOF}. Figure \ref{F:nHe3setup} shows the basic scheme. A 0.235 nm wavelength neutron beam was extracted and focused by two pyrolytic graphite monochromators. A graphite beam filter reduced contamination from higher order Bragg reflections (to $<10^{-3}$). The beam then passed through a transmission-mode supermirror in which one spin state was preferentially reflected and absorbed and the other transmitted, producing 93\% neutron polarization. Following the polarizer was a spin flipper that rotated the neutron spin by 180$^{\circ}$ by Larmor precession when energized. A pair of 56-cm diameter Helmholtz coils produced a holding field of 1.48 mT centered on the $^3$He cell. The perfect crystal silicon interferometer was the skew-symmetric type with (220) reflecting planes. The initial (no target) fringe contrast was 85\%.
\par
A novel feature of this experiment was the use of a polarized gas target in the neutron interferometer. The desired precision and the space available inside the interferometer called for a small target cell with uniform length and high polarization. Space constraints as well as the need to avoid thermal gradients near the interferometer precluded polarizing the cell {\em in situ}. Therefore we required a cell with a long spin-relaxation time so it could be polarized outside the apparatus, transported and positioned in the interferometer, and remain polarized in place for several days of measurement. We used the spin-exchange optical pumping (SEOP) method (see {\em e.g.} \cite{Wal97,Ric02}) to polarize the target. Two different target cells, with {\em in situ} spin relaxation times of 115 h and 35 h, were used in the experiment. The intrinsic relaxation time of the longer-lived cell was measured to be 370 h \cite{Bab06}. The cells were cylindrical with optically sealed 4 mm-thick flat windows, fabricated from boron-free aluminosilicate glass \cite{Gen07}. The outside dimensions were 42 mm length and 25 mm diameter. Each cell was filled to a $^3$He pressure of approximately 180 kPa (1.8 bar), chosen to roughly minimize the statistical uncertainty. When a maximum $^3$He polarization of about 65\% was obtained in the polarization setup, the cell was moved to the neutron interferometer.
\par
\begin{figure}
\includegraphics[width=5.0in]{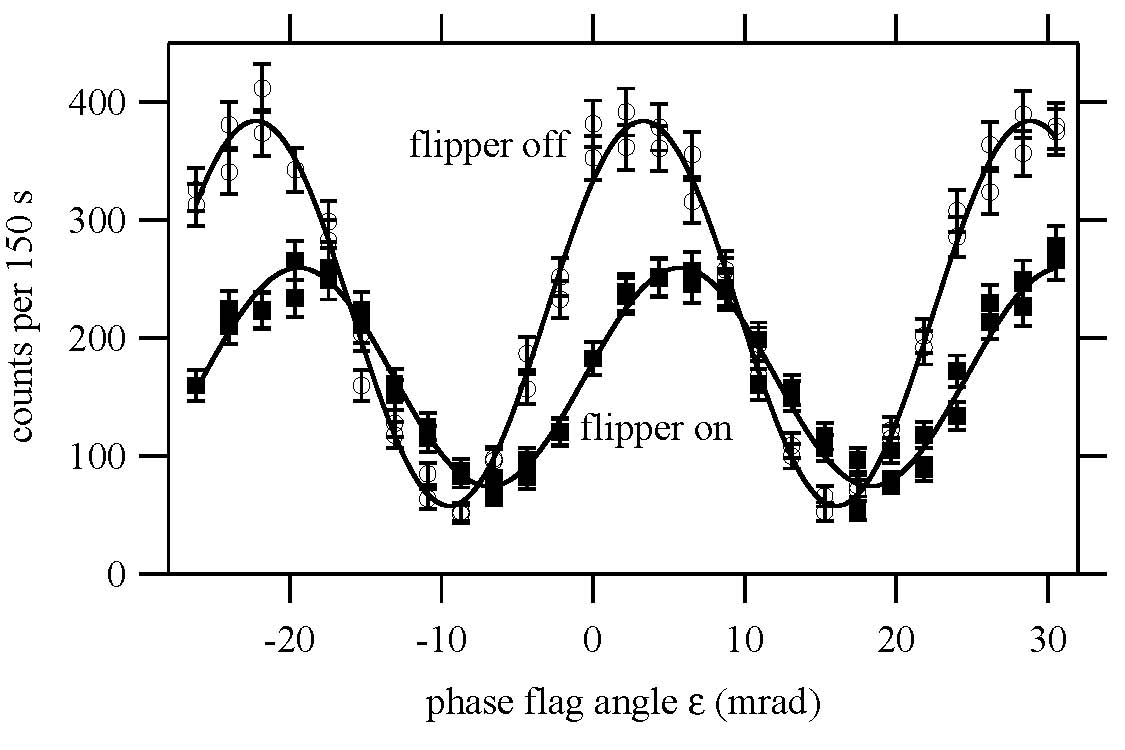}
\caption{\label{F:scan}A typical phase scan, showing counts in the O-beam detector (C3) for neutron spin parallel (open circles) and
antiparallel (solid squares) to the $^3$He polarization, and their best fits to equation \ref{E:cosfcn}. }
\end{figure}

The data were collected in scans that lasted 4 h -- 9 h each. For each scan the phase flag angle $\epsilon$ -- relative to the plane of the interferometer blades -- was moved over a small range in steps of 2.18 mrad. At each phase flag position the spin flipper was modulated in an off-on-on-off sequence and the count rate in the O-beam detector (C3 in figure \ref{F:nHe3setup}) was recorded. A typical scan is shown in figure \ref{F:scan}. The count rate for each spin flipper state was fit to
\begin{equation}
\label{E:cosfcn}
I(\epsilon) = c_0 + c_1 \cos \left( c_2 f(\epsilon) + \phi \right)
\end{equation}
where $c_0$, $c_1$, $c_2$, and $\phi$ are fit parameters and
\begin{equation}
f(\epsilon) = \frac{\sin\theta_{\rm B} \sin\epsilon}{\cos^2\theta_{\rm B} - \sin^2\epsilon}
\end{equation}
is the geometrical factor for the phase flag, with $\theta_{\rm B}$ the Bragg angle of the interferometer. The difference in the extracted $\phi$ for the two fits is $\Delta\phi = (\phi^\uparrow - \phi^\downarrow)$. Typically the reduced $\chi^2$ for these fits were between 0.8 and 1.3. Poor quality scans, in which the reduced $\chi^2 > 1.5$, were discarded. Most of the poor scans were due to phase instability immediately following a cell transfer, which caused temperature and mechanical fluctuations that persisted for several hours.
\par
The product $N_3 \lambda z P_3$ needed in equation \ref{E:maineq} was measured continuously during the experiment. The first three factors were constant but $P_3$ decreased exponentially with time as the cell polarization relaxed. The stronger absorption for one neutron spin state causes an asymmetry $A_4$, that depends on $P_3$, between the count rates $N^\uparrow$ and $N^\downarrow$ measured in detector C4 (see figure \ref{F:nHe3setup})
\begin{equation}
A_4 = \frac{N^{\uparrow} - N^{\downarrow}}{N^{\uparrow} + N^{\downarrow}} = 
\frac{\frac{1}{2} (1+s) P_n \tanh\xi}{1 + \frac{1}{2} (1-s) P_n \tanh\xi}
\end{equation}
with
\begin{equation}
\label{E:xi}
\xi = \left(\frac{\sigma_0 - \sigma_1}{4 \lambda_{\rm th}}\right) N_3 \lambda z P_3,
\end{equation}
where $\sigma_0$ ($\sigma_1$) is the $^3$He thermal neutron absorption cross section for the singlet (triplet) channel and $\lambda_{\rm th}$ is the reference thermal neutron wavelength (0.1798 nm). The total thermal absorption cross section $\sigma_{\rm th} = \frac{1}{4}\sigma_0 + \frac{3}{4}\sigma_1$ = 5333(7) b \cite{3Hexs} is dominated by $\sigma_0$ due to the strong singlet resonance at -0.5 MeV, but the triplet cross section is not well known \cite{3Hexs1}. In the literature it has often been assumed that $\sigma_1 = 0$  but  as pointed out in \cite{Zim02}  that assumption is not justified. Hofmann and Hale recently calculated the imaginary triplet scattering length $a_1''$ using the $R$-matrix and AV18 with 3N forces and obtained $-0.001$ fm and $-0.005$ fm, respectively \cite{Hof03}. In the same paper they show that AV18+3N underpredicts $a_0''$ by as much as 30\% compared to the experimental value. 
Therefore to be conservative we use $a_1'' = (-0.005\pm0.005)$ fm; this range includes the $R$-matrix value and allows for an underestimate by a factor of two by AV18+3N. With $\sigma_1 / \sigma_0 = a_1''/a_0''$ we have $\sigma_0 - \sigma_1$ = (21,236 $\pm$ 28 experimental $\pm$ 100 theoretical) b.
The average value of $A_4$ for one scan was used to extract $N_3 \lambda z P_3$ for that scan. Studies using simulated data verified that the systematic error due to using the average value of $A_4$ was negligible.
\par
Because the neutron polarization is not 100\% the interferogram (Eq. \ref{E:cosfcn}) is actually a sum of two interferograms, one for each neutron spin state, with relative intensities given by the ratio of neutron spin states at the exit of the $^3$He cell. We apply a correction to the phase difference between the two spin flipper states:
\begin{equation}
\label{E:deltaphi}
\Delta\phi_{\rm meas} = \arctan\left( \frac{\sin\Delta\phi}{\eta^\downarrow + \cos\Delta\phi} \right) -  
\arctan\left( \frac{\eta^\uparrow\sin\Delta\phi}{1 + \eta^\uparrow\cos\Delta\phi} \right).
\end{equation}
Here $\Delta\phi_{\rm meas}$ is the measured phase difference, $\Delta\phi$ is the phase difference one would obtain with 100\% neutron polarization, 
and $\eta^\uparrow$ ($\eta^\downarrow$) is the ratio of minority to majority neutron spin states at the exit of the cell with the spin flipper off (on), given by
\begin{equation}
\label{E:etaup}
\eta^\uparrow = \left( \frac{1 - P_n}{1 + P_n} \right) e^{-2 \xi}
\end{equation}
\begin{equation}
\label{E:etadown}
\eta^\downarrow= \left( \frac{1 - sP_n}{1 + sP_n} \right) e^{+2 \xi}.
\end{equation}
$P_n$ is the neutron polarization and $s$ accounts for the slight decrease in polarization with the spin flipper energized,
{\em i.e.} $P_n\equiv  P_n^\uparrow$ and $s = P_n^\downarrow / P_n^\uparrow$. 
\par
The neutron polarization at the interferometer entrance, for both spin flipper states, was determined 16 times during the six month duration of the experiment by replacing the interferometer crystal with an optically thick ($N_3\sigma z\approx 3$) polarized $^3$He analyzer cell. We obtained $P_n = 0.9291\pm0.0008$ and $s = 0.9951\pm0.0003$. This was stable over the course of the experiment.
\par
A potential systematic error can arise because the magnetic field in the interferometer is not perfectly uniform. A field gradient will cause a phase difference between the two neutron paths that reverses sign, so does not cancel, when the spin flipper is energized. The size of this effect was determined in three ways: 1) direct measurement by repeating the experiment with the cell removed, 2) extrapolation of the data to $P_3$ = 0, and 3) an estimate of the gradient from the observed relaxation time of $P_3$. The first two methods were consistent with zero. The third indicated a small effect with an associated correction to $\Delta b'$ of 0.009 fm. Since we don't know the sign we take this to be the systematic uncertainty due to the magnetic field gradient and make no correction.
\par
\begin{figure}
\includegraphics[width=6.0in]{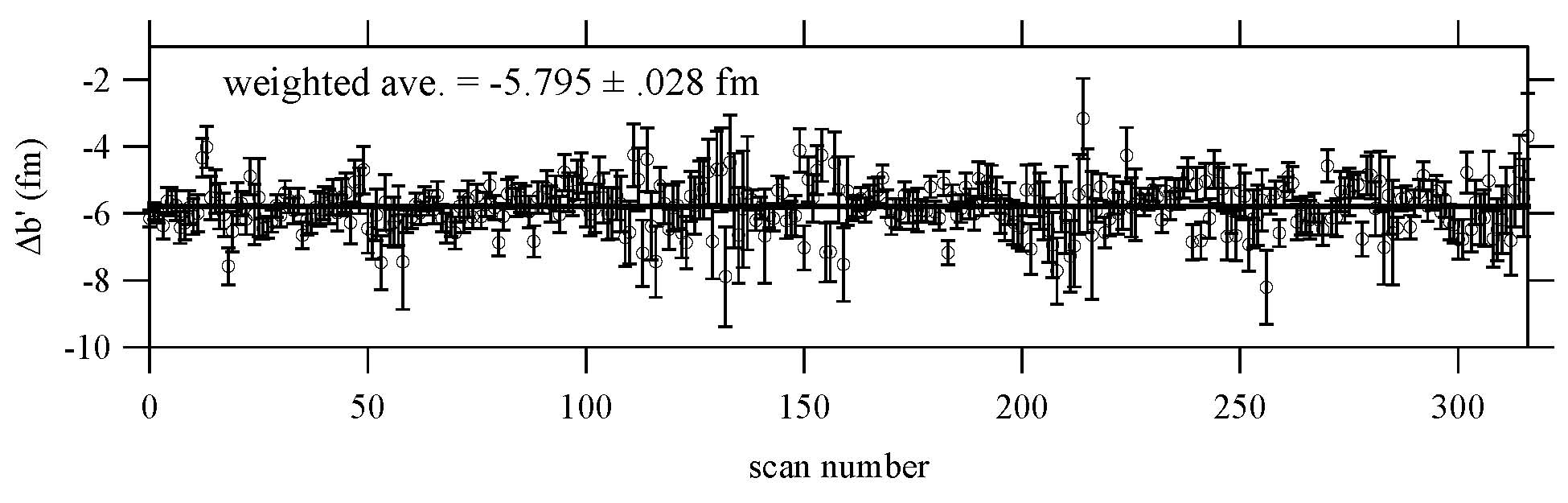}
\caption{\label{F:allscans} $\Delta b'$ for 317 scans. Error bars shown are counting statistics only. The weighted average is ($-5.795\pm.028$) fm with 
reduced $\chi^2$ = 371/316.}
\end{figure}
\begin{figure}
\includegraphics[width=5.0in]{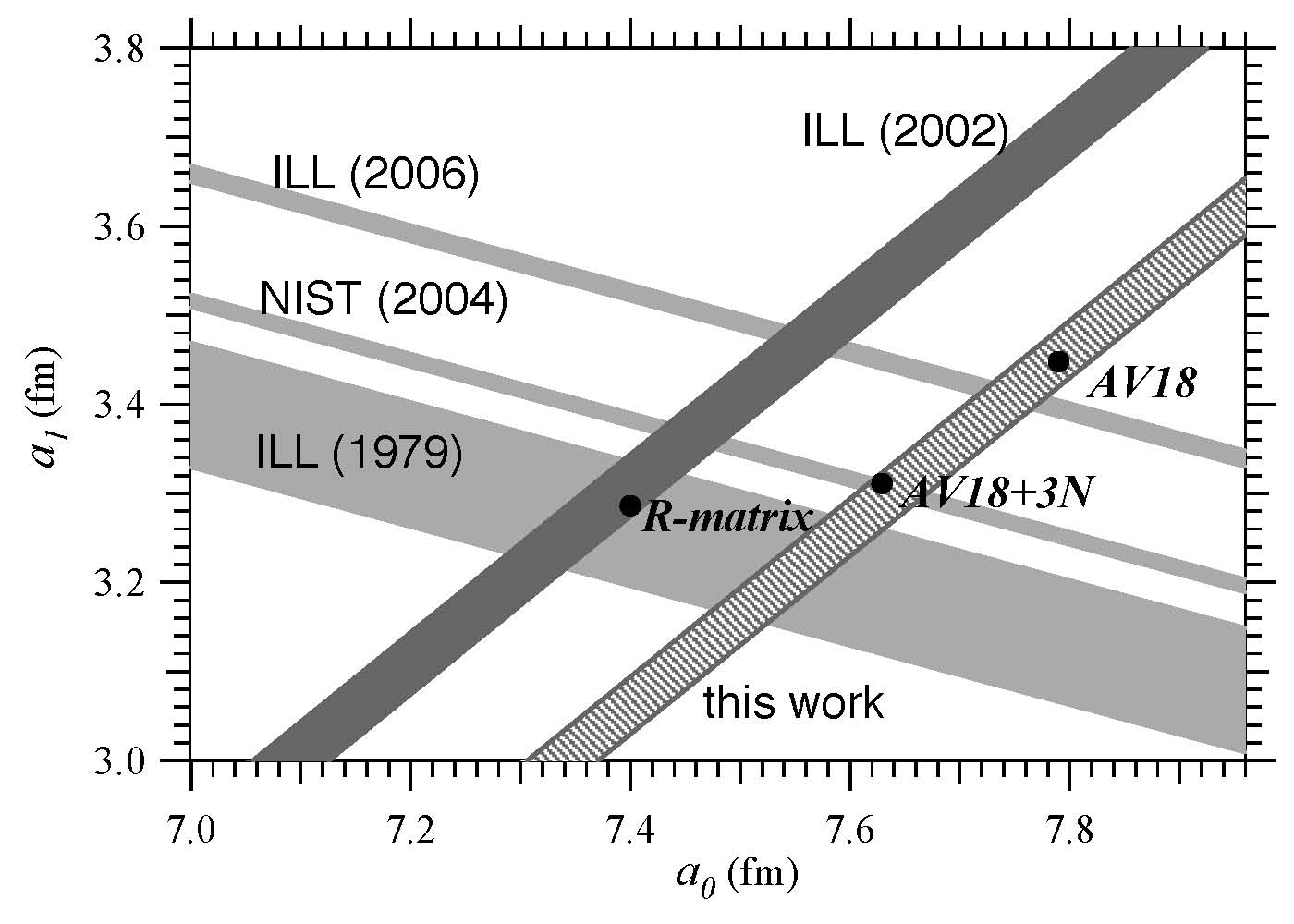}
\caption{\label{F:nHe3sum} Summary of the triplet and singlet $n$-$^3$He free neutron scattering lengths (real parts).
The three coherent (unpolarized) measurements labelled ILL (1979) \cite{Kai79}, NIST (2004) \cite{Huf04}, and ILL (2006) \cite{Ket06}
used neutron interferometry. The incoherent (polarized) measurement ILL (2002) \cite{Zim02} used pseudomagnetic spin
rotation. The three theoretical points shown are from \cite{Hof03}.}
\end{figure}

For each scan we numerically inverted Eq. \ref{E:deltaphi} to extract $\Delta\phi$ from $\Delta\phi_{\rm meas}$ and then divided by 
$N_3 \lambda z P_3/2$ to obtain a value for $\Delta b'$ (see Eq. \ref{E:maineq}). Figure \ref{F:allscans} shows $\Delta b'$ for 317 scans from 12 weeks of measurements. The weighted average is $(-5.795\pm0.028)$ fm with reduced $\chi^2$= 371/316. The low probability of this fit (about 2\%) indicates a small amount of phase instability, mainly due to temperature fluctuations, that is typical for this instrument. Because the instability is quasirandom, we add in quadrature an additional uncertainty of 0.20 fm to each scan in order to obtain reduced $\chi^2$ = 1. The new weighted average is $(-5.802\pm0.031)$ fm, hence we obtain
\begin{equation}
\Delta b' = (-5.802\pm0.028\mbox{ stat.}\pm0.033\mbox{ sys.})\mbox{ fm}
\end{equation}
using the combined systematic uncertainty from Table \ref{T:Errors}. With equation \ref{E:incoh} we can express this result as the real part of the incoherent scattering length: $b_i' = (-2.512\pm0.012\mbox{ stat.}\pm0.014\mbox{ sys.})$ fm. In terms of the free scattering length it is 
$a_1' - a_0' = (-4.346\pm0.021\mbox{ stat.}\pm0.025\mbox{ sys.})$ fm. The poorly-known ratio $\sigma_1 / \sigma_0$ contributes the largest systematic uncertainty both to this experiment and Ref.~\cite{Zim02}. It may be improved in future so it is useful to express our result with this quantity factored out:
$b_i' = (-2.515 (1-\sigma_1/\sigma_0)\pm0.012\mbox{ stat.}\pm0.008\mbox{ sys.})$ fm.
Figure \ref{F:nHe3sum} shows a summary of $a_0$ and $a_1$ in the $n$-$^3$He system from the most recent coherent and incoherent scattering length measurements. Our result is in good agreement with the calculation by Hoffman and Hale using the AV18+3N potential \cite{Hof03}, but disagrees with the only previous measurement of this quantity by Zimmer {\em et al.}, $b_i' = (-2.365\pm 0.020)$ fm \cite{Zim02}, which used a very different technique, pseudomagnetic spin rotation. The disagreement between the various experiments remains a problem that must be resolved by future work.

\begin{table}
\caption{\label{T:Errors} Uncertainty budget for $\Delta b'$. All uncertainties quoted in this paper are one sigma.}
\begin{ruledtabular}
\begin{tabular}{p{2.0in}cc}
Source 						&Uncertainty (fm)\\
\hline
phase instability				&0.013\\
neutron polarization $P_n$		&0.006\\
neutron spin flip factor $s$		&0.002\\
magnetic field gradient                        &0.009\\
$n$-$^3$He thermal abs. cross section $\sigma_{\rm th}$	&0.008\\
$n$-$^3$He triplet abs. cross section $\sigma_1$			&0.027\\
\hline
combined systematic uncertainty		&0.033\\
counting statistics				&0.028\\
\end{tabular}
\end{ruledtabular}
\end{table}
\par
We are grateful to the NIST Center for Neutron Research for providing the neutron beam and technical support. We thank Sam Werner and Helmut Kaiser for helpful discussions This work was supported by NIST (U.~S.~Dept. of Commerce) and the National Science Foundation through grant PHY-0555347. The development and application of the polarized $^3$He cells was supported in part by the U.~S.~Dept. of Energy, Basic Energy Sciences.

\end{document}